\documentclass[showpacs,aps,prd]{revtex4}
\usepackage{CJK,fancyhdr}
\input epsf

\begin{document}

\title{$0^{+}$ tetraquark states from improved QCD sum rules: delving into $X(5568)$}
\author{Jian-Rong Zhang ÕŽ¨ÈÙ}
\affiliation{Department of Physics, College of Liberal Arts and Sciences, National University of Defense Technology,
Changsha 410073, Hunan, People's Republic of China}
\author{Jing-Lan Zou ×Þ¾°á°}
\affiliation{College of Optoelectronic Science and Engineering, National University of Defense Technology,
Changsha 410073, Hunan, People's Republic of China}
\author{Jin-Yun Wu Îâ½ðÔÆ}
\affiliation{College of Liberal Arts and Sciences, National University of Defense Technology,
Changsha 410073, Hunan, People's Republic of China}


\begin{abstract}
In order to investigate the possibility of the recently observed $X(5568)$ being a $0^{+}$ tetraquark state,
we make an improvement to the study of the related various configuration states
in the framework of the QCD sum rules.
Particularly, to ensure the quality of the analysis, condensates
up to dimension $12$ are included to inspect the convergence of operator product expansion (OPE) and
improve the final results of the studied states.
We note that some condensate contributions could play an important role on the OPE side.
By releasing the rigid
OPE convergence criterion,
we arrive at the
numerical value $5.57^{+0.35}_{-0.23}~\mbox{GeV}$ for the
 scalar-scalar diquark-antidiquark $0^{+}$ state,
which agrees with the experimental data for the $X(5568)$ and could support
its interpretation
in terms of a $0^{+}$ tetraquark state with the
 scalar-scalar configuration.
The corresponding result for the axial-axial current
is calculated to be $5.77^{+0.44}_{-0.33}~\mbox{GeV}$,
which is still consistent with the mass of $X(5568)$ in view of the uncertainty.
The feasibility of $X(5568)$ being a tetraquark state with the axial-axial configuration therefore cannot be definitely excluded.
For the pseudoscalar-pseudoscalar and the vector-vector cases, their
 unsatisfactory OPE convergence make it difficult to find
reasonable work windows to extract the hadronic information.
\end{abstract}
\maketitle

\section{Introduction}\label{sec1}
Not long ago, the D0 Collaboration reported evidence for a narrow structure, referred to
as the $X(5568)$, in the decay modes
$X(5568)\rightarrow B_{s}^{0}\pi^{\pm}$ produced in $p\bar{p}$ collisions at center-of-mass energy
$\sqrt{s}=1.96~\mbox{TeV}$ \cite{X5568}. Its mass and natural width were measured
to be $m=5567.8\pm2.9_{-1.9}^{+0.9}~\mbox{MeV}$ and $\Gamma=21.9\pm6.4_{-2.5}^{+5.0}~\mbox{MeV}$, respectively.
With the $B_{s}^{0}\pi^{+}$
produced in an $S$-wave, its quantum
number would be $J^{P}=0^{+}$.
Subsequently, the LHCb Collaboration announced that the existence of
$X(5568)$ was not confirmed
in the analysis of
$pp$ collision at energies $7~\mbox{TeV}$ and $8~\mbox{TeV}$ \cite{X5568-1},
and the CMS Collaboration did not find the $X(5568)$ structure \cite{X5568-2} either.
However, the D0 Collaboration then observed the $X(5568)$ again
in the $B_{s}^{0}\pi^{\pm}$
invariant mass distribution via another channel
 $B_{s}^{0}\rightarrow D_{s}\mu\nu$ at the
same mass and at the expected width and rate \cite{X5568-3}.
One explanation for the $X(5568)$ appearance
in D0 and its absence in LHCb and CMS was proposed in Ref. \cite{X5568-4}.

The $X(5568)$ has not only attracted experimental attention,
but also aroused great enthusiasm from theorists
in attempting to
understand its underlying structure \cite{X5568-Tetra,X5568-Th} (for
recent reviews, see e.g. Refs.~\cite{X5568-rev,X5568-rev1} and references
therein).
As an imaginable scenario,
a $0^{+}$
tetraquark state with four different valence quark flavors has been proposed as a potential candidate \cite{X5568,X5568-Tetra}.
Without doubt, it is important to investigate whether $X(5568)$
can be interpreted as a tetraquark state,
which could provide a crucial
piece of information to help understand how exotic hadrons
are bound, and comprehend QCD more deeply at low energy.
However, it is difficult to quantitatively acquire the hadronic information,
in view of our limited understanding of QCD's nonperturbative aspects.

The QCD sum rule approach \cite{svzsum} is a nonperturbative
formulation firmly grounded on the QCD theory,
which
has already been widely and successfully applied
to research many hadrons \cite{overview1,overview2,overview3,overview4}.
With regard to the recently observed $X(5568)$,
there have been several studies using the QCD sum rules
to study its mass from the point of view of a $0^{+}$ tetraquark state \cite{Azizi,Wang,Zhu,Nielsen,Qiao,Narison},
chiefly focusing on some particular configurations.
Firstly, one can employ various configurations, e.g. scalar-scalar,
pseudoscalar-pseudoscalar, axial vector-axial vector (shortened to ``axial-axial'' below), and  vector-vector
diquark-antidiquark, to construct a $0^{+}$ tetraquark current and work over these possible configurations.
Secondly,
for the QCD sum rule method, one of its key points is that
 both the OPE convergence and
the pole dominance should be meticulously inspected to determine the work window,
ensuring the credibility of the obtained result.
It may be difficult to satisfy both the above criteria, because in some cases
it may be hard to find a work window critically satisfying
both rules. This could become specially obvious for some multiquark states (e.g. see discussions in Refs.~\cite{Zs0,Zs1,Zs2,Zs}).
The main reason is that some high dimension condensates
may play an important role on the OPE
side, which means that the standard OPE convergence may happen only at large values of the Borel parameters.
Therefore, it may be more reliable to test the OPE convergence by taking into account
higher dimension condensates and fixing
the work windows precisely. One can then
obtain the hadronic properties more
safely.
Even if higher condensates
do not radically influence the character of OPE convergence in some cases, to say the least,
one still could expect to improve the final result, since
higher dimensional condensates are helpful to stabilize the Borel curve.
In order to uncover the inner structure of $X(5568)$,
it is significant and worthwhile to make further theoretical
efforts. From the above two considerations,  we
endeavor to perform an improved sum rule study
on whether $X(5568)$ could be a $0^{+}$ tetraquark state.
In particular, we carry out calculations with four different configuration currents and
pay close attention to higher dimension condensate effects.

The rest of this paper is organized as follows. In Section \ref{sec2}, QCD sum
rules for the tetraquark states are derived, involving both the
phenomenological representation and the QCD side, which is followed
by  numerical analysis and some discussion in Section \ref{sec3}.
The last section give a brief summary.

\section{Tetraquark state QCD sum rules}\label{sec2}
In the QCD sum rules, one basic point is to build a proper
interpolating current to represent the studied state.
For a tetraquark state, its current could be constructed as
the usual diquark-antidiquark configuration.
Hence, one can obtain the following form of current:
\begin{eqnarray}
j_{(i)}&=&\epsilon_{abc}\epsilon_{dec}(q_{a}^{T}\Gamma_{i}s_{b})(\bar{q}_{d}\Gamma'_{i}\bar{Q}_{e}^{T})\nonumber
\end{eqnarray}
for the tetraquark sate,
where the index $i$ takes $I,II,III$, or $IV$, $q$ indicates the light $u$ or $d$ quark, $Q$ denotes the heavy quark, and the subscripts $a$,
$b$, $c$, $d$, and $e$ are color indices.
To form currents with a total quantum number $J^{P}=0^{+}$,
$\Gamma$ matrices are taken as $\Gamma_{I}=C\gamma_{5}$,  $\Gamma'_{I}=\gamma_{5}C$
for the scalar-scalar case,
$\Gamma_{II}=C$, $\Gamma'_{II}=C$ for the pseudoscalar-pseudoscalar case,
$\Gamma_{III}=C\gamma_{\mu}$,  $\Gamma'_{III}=\gamma^{\mu}C$ for the axial-axial case,
and $\Gamma_{IV}=C\gamma_{5}\gamma_{\mu}$, $\Gamma'_{IV}=\gamma^{\mu}\gamma_{5}C$ for the vector-vector case.

Further, the two-point correlator,
\begin{eqnarray}
\Pi_{i}(q^{2})=i\int
d^{4}x\mbox{e}^{iq.x}\langle0|T[j_{(i)}(x)j_{(i)}^{\dag}(0)]|0\rangle,
\end{eqnarray}
can be used to derive the tetraquark state QCD sum rules.

The correlator
$\Pi_{i}(q^{2})$ can be phenomenologically expressed as
\begin{eqnarray}\label{ph}
\Pi_{i}(q^{2})=\frac{\lambda_{H}^{2}}{M_{H}^{2}-q^{2}}+\frac{1}{\pi}\int_{s_{0}}
^{\infty}\frac{\mbox{Im}\big[\Pi_{i}^{\mbox{phen}}(s)\big]}{s-q^{2}}ds+...,
\end{eqnarray}
where $M_{H}$ is the mass of the hadronic state,
$s_0$ denotes the continuum threshold, and $\lambda_{H}$ indicates
the coupling of the current to the hadron $\langle0|j|H\rangle=\lambda_{H}$.

On the OPE side, the correlator $\Pi_{i}(q^{2})$ can be theoretically written as
\begin{eqnarray}\label{ope}
\Pi_{i}(q^{2})=\int_{(m_{Q}+m_{s})^{2}}^{\infty}\frac{\rho_{i}(s)}{s-q^{2}}ds+\Pi_{i}^{\mbox{cond}}(q^{2}),
\end{eqnarray}
where the spectral density $\rho_{i}(s)$ is
$\frac{1}{\pi}\mbox{Im}\big[\Pi_{i}(s)\big]$, $m_{Q}$ is the heavy quark mass, and $m_{s}$ is the strange quark mass.
In the concrete derivation, one can work at leading order
in $\alpha_{s}$ and take into account condensates up to dimension $12$,
with similar techniques as in Refs. \cite{overview4,Tech,Zhang}. To keep the heavy-quark mass $m_{Q}$ finite, one can use the heavy-quark propagator
in the momentum space \cite{reinders},
\begin{eqnarray}
S_{Q}(p)&=&\frac{i}{\rlap/p-m_{Q}}
-\frac{i}{4}gt^{A}G^{A}_{\kappa\lambda}(0)\frac{1}{(p^{2}-m_{Q}^{2})^{2}}\Big[\sigma^{\kappa\lambda}(\rlap/p+m_{Q})
+(\rlap/p+m_{Q})\sigma^{\kappa\lambda}\Big]\nonumber\\&&{}
-\frac{i}{4}g^{2}t^{A}t^{B}G^{A}_{\alpha\beta}(0)G^{B}_{\mu\nu}(0)\frac{\rlap/p+m_{Q}}{(p^{2}-m_{Q}^{2})^{5}}\Big[
\gamma^{\alpha}(\rlap/p+m_{Q})\gamma^{\beta}(\rlap/p+m_{Q})\gamma^{\mu}(\rlap/p+m_{Q})\gamma^{\nu}\\&&{}
+\gamma^{\alpha}(\rlap/p+m_{Q})\gamma^{\mu}(\rlap/p+m_{Q})\gamma^{\beta}(\rlap/p+m_{Q})\gamma^{\nu}+
\gamma^{\alpha}(\rlap/p+m_{Q})\gamma^{\mu}(\rlap/p+m_{Q})\gamma^{\nu}(\rlap/p+m_{Q})\gamma^{\beta}\Big](\rlap/p+m_{Q})\nonumber\\&&{}
+\frac{i}{48}g^{3}f^{ABC}G^{A}_{\gamma\delta}G^{B}_{\delta\varepsilon}G^{C}_{\varepsilon\gamma}\frac{1}{(p^{2}-m_{Q}^{2})^{6}}(\rlap/p+m_{Q})
\Big[\rlap/p(p^{2}-3m_{Q}^{2})+2m_{Q}(2p^{2}-m_{Q}^{2})\Big](\rlap/p+m_{Q}).\nonumber
\end{eqnarray}
The light-quark part of the
correlator can be calculated in the coordinate space, with the light-quark
propagator,
\begin{eqnarray}
S_{ab}(x)&=&\frac{i\delta_{ab}}{2\pi^{2}x^{4}}\rlap/x-\frac{m_{q}\delta_{ab}}{4\pi^{2}x^{2}}-\frac{i}{32\pi^{2}x^{2}}t^{A}_{ab}gG^{A}_{\mu\nu}(\rlap/x\sigma^{\mu\nu}
+\sigma^{\mu\nu}\rlap/x)-\frac{\delta_{ab}}{12}\langle\bar{q}q\rangle+\frac{i\delta_{ab}}{48}m_{q}\langle\bar{q}q\rangle\rlap/x\nonumber\\&&{}\hspace{-0.3cm}
-\frac{x^{2}\delta_{ab}}{3\cdot2^{6}}\langle g\bar{q}\sigma\cdot Gq\rangle
+\frac{ix^{2}\delta_{ab}}{2^{7}\cdot3^{2}}m_{q}\langle g\bar{q}\sigma\cdot Gq\rangle\rlap/x-\frac{x^{4}\delta_{ab}}{2^{10}\cdot3^{3}}\langle\bar{q}q\rangle\langle g^{2}G^{2}\rangle,
\end{eqnarray}
which is then
Fourier-transformed to the momentum space in $D$ dimensions.
The strange quark is treated as a light one and the diagrams are
considered up to order $m_{s}$.
The
resulting light-quark part is combined with the heavy-quark part
before it is dimensionally regularized at $D=4$.
After equating Eqs. (\ref{ph}) and (\ref{ope}), utilizing quark-hadron duality, and
doing a Borel transform $\hat{B}$, the sum rule can be
\begin{eqnarray}\label{sumrule1}
\lambda_{H}^{2}e^{-M_{H}^{2}/M^{2}}&=&\int_{(m_{Q}+m_{s})^{2}}^{s_{0}}\rho_{i}(s)e^{-s/M^{2}}ds+\hat{B}\Pi_{i}^{\mbox{cond}},
\end{eqnarray}
with $M^2$ the Borel parameter.
For compactness, the concrete forms of $\rho_{i}(s)$
and $\hat{B}\Pi_{i}^{\mbox{cond}}$
are shown in the Appendix.

Taking
the derivative of the sum rule (\ref{sumrule1}) in terms of $-\frac{1}{M^2}$ and then dividing
by itself, one can get the mass of the hadronic state
\begin{eqnarray}\label{sum rule 1}
M_{H}&=&\sqrt{\bigg\{\int_{(m_{Q}+m_{s})^{2}}^{s_{0}}\rho_{i}(s)s
e^{-s/M^{2}}ds+\frac{d\big(\hat{B}\Pi_{i}^{\mbox{cond}}\big)}{d(-\frac{1}{M^{2}})}\bigg\}/
\bigg\{\int_{(m_{Q}+m_{s})^{2}}^{s_{0}}\rho_{i}(s)e^{-s/M^{2}}ds+\hat{B}\Pi_{i}^{\mbox{cond}}\bigg\}},
\end{eqnarray}
with $i=I$, $II$, $III$, or $IV$.

\section{Numerical analysis and discussion}\label{sec3}
In this section, we perform numerical analysis of the sum rule (\ref{sum rule 1})
to extract the mass value of the studied state. The input parameters
are taken as $m_{b}=4.18^{+0.04}_{-0.03}~\mbox{GeV}$,
$m_{s}=96^{+8}_{-4}~\mbox{MeV}$,
$\langle\bar{q}q\rangle=-(0.24\pm0.01)^{3}~\mbox{GeV}^{3}$,
$\langle\bar{s}s\rangle=m_{0}^{2}~\langle\bar{q}q\rangle$,
$\langle
g\bar{q}\sigma\cdot G q\rangle=m_{0}^{2}~\langle\bar{q}q\rangle$,
$m_{0}^{2}=0.8\pm0.1~\mbox{GeV}^{2}$, $\langle
g^{2}G^{2}\rangle=0.88\pm0.25~\mbox{GeV}^{4}$, and $\langle
g^{3}G^{3}\rangle=0.58\pm0.18~\mbox{GeV}^{6}$ \cite{PDG,svzsum,overview2}.
In the standard
procedure,
one should consider both the OPE convergence and the pole contribution dominance
to choose
proper work windows for the threshold $\sqrt{s_{0}}$ and the Borel
parameter $M^{2}$.
At the same time, the threshold
$\sqrt{s_{0}}$ cannot be taken optionally.
This is because $\sqrt{s_{0}}$ characterizes the
beginning of continuum states.
In practice, it may be difficult to find a
conventional work window that critically satisfies all the
rules in some studies (for instance, see Refs.~\cite{Zs0,Zs1,Zs2,Zs}).

One can illustrate the numerical analysis process
by giving the scalar-scalar case as an example.
Its various contributions
are compared as a function of $M^2$ and shown
in Fig.~1 to test the
convergence of OPE.
There
are three main condensate contributions, i.e. the two-quark condensate,
the four-quark condensate, and the mixed condensate.
These condensates could play an important role on the OPE side,
which makes the standard OPE convergence happen only at very large values of $M^2$.
The consequence is that it is difficult to find a conventional Borel window
strictly satisfying both that the pole dominates over the continuum and the OPE converges well.
It is not too bad for the present case: there are three main condensates
and they could cancel each other out
to some extent, as they have different signs.
What is also very important is that most
of other condensates calculated are very small,
almost negligible,
which means that they cannot radically influence the character of OPE convergence.
All these factors mean that
one could find the perturbative dominance in the total
and the OPE convergence still be under control.
Without any adverse consequences, one could try releasing the rigid
OPE convergence criterion for the present case and
choose proper work windows at relatively
low values of $M^{2}$.

On the phenomenological side,
the comparison between pole and
continuum contributions of sum rule (\ref{sumrule1}) as a function of the
Borel parameter $M^2$ for the threshold value
$\sqrt{s_{0}}=6.1~\mbox{GeV}$ is shown in Fig.~2, which shows that the relative pole
contribution is approximate to $50\%$ at $M^{2}=3.8~\mbox{GeV}^{2}$
and decreases with $M^{2}$. Similarly, the upper bound values of the Borel parameters are
$M^{2}=3.7~\mbox{GeV}^{2}$ for
$\sqrt{s_0}=6.0~\mbox{GeV}$ and $M^{2}=3.9~\mbox{GeV}^{2}$ for
$\sqrt{s_0}=6.2~\mbox{GeV}$.
Therefore, the Borel windows for
the scalar-scalar diquark-antidiquark state are taken as
$2.7\sim3.7~\mbox{GeV}^{2}$ for $\sqrt{s_0}=6.0~\mbox{GeV}$,
$2.7\sim3.8~\mbox{GeV}^{2}$ for
$\sqrt{s_0}=6.1~\mbox{GeV}$, and
$2.7\sim3.9~\mbox{GeV}^{2}$ for $\sqrt{s_0}=6.2~\mbox{GeV}$. The mass of the $0^{+}$ tetraquark state
with the scalar-scalar configuration as
a function of $M^2$ from sum rule (\ref{sum rule 1}) is
shown in Fig.~3 and it is numerically counted to be $5.57\pm0.19~\mbox{GeV}$ in the chosen work windows.
Considering the uncertainty from the variation of quark masses and
condensates, we get
$5.57\pm0.19^{+0.16}_{-0.04}~\mbox{GeV}$ (the
first error reflects the uncertainty due to variation of $\sqrt{s_{0}}$
and $M^{2}$, and the second error results from the variation of
QCD parameters) or concisely $5.57^{+0.35}_{-0.23}~\mbox{GeV}$
for the scalar-scalar tetraquark state.

\begin{figure}[htb!]
\centerline{\epsfysize=5.18truecm\epsfbox{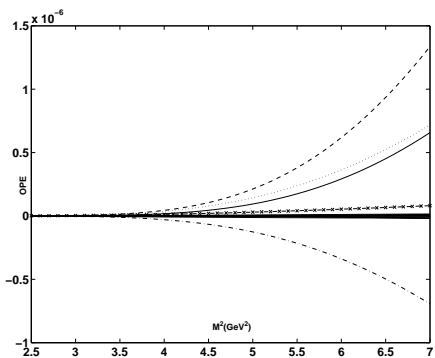}}
\caption{The various OPE contribution as a function of $M^2$ in sum rule
(\ref{sumrule1}) for $\sqrt{s_{0}}=6.1~\mbox{GeV}$ for the scalar-scalar case.
Four main contributions, i.e. the perturbative, the two-quark condensate,
the four-quark condensate, and the mixed condensate
are denoted by the single solid line,
the dashed line, the dotted line, and the dot-dashed line, respectively.
These main condensates could cancel each other out
to some extent, and
other condensates are very small.
All these factors mean that one can find perturbative dominance in the total
and it is still under control for OPE convergence.}
\end{figure}

\begin{figure}
\centerline{\epsfysize=5.18truecm\epsfbox{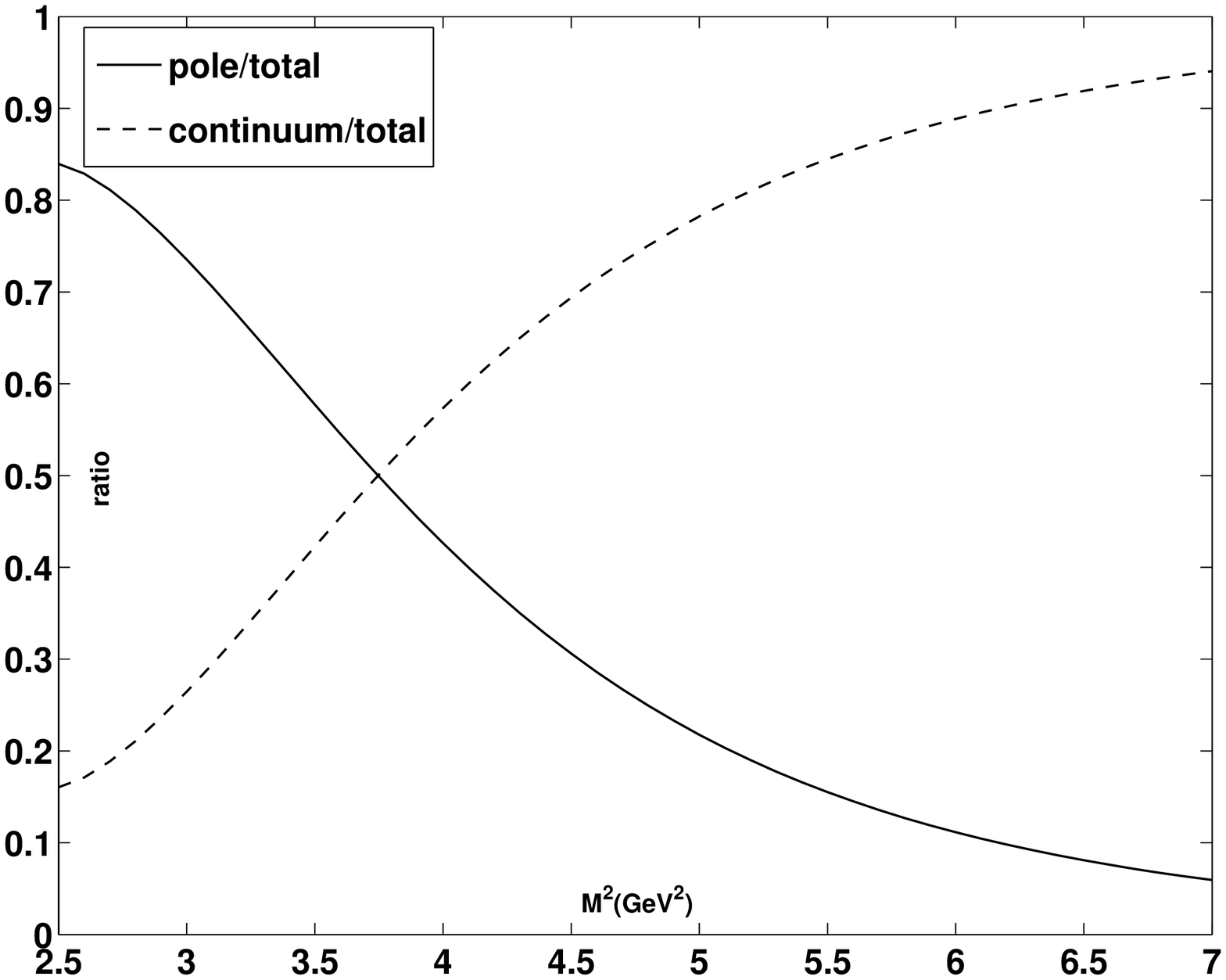}}
\caption{The phenomenological contribution in sum rule
(\ref{sumrule1}) for $\sqrt{s_{0}}=6.1~\mbox{GeV}$ for the scalar-scalar case.
The solid line is the relative pole contribution (the pole
contribution divided by the total, pole plus continuum contribution)
as a function of $M^2$ and the dashed line is the relative continuum
contribution.}
\end{figure}

\begin{figure}
\centerline{\epsfysize=5.18truecm
\epsfbox{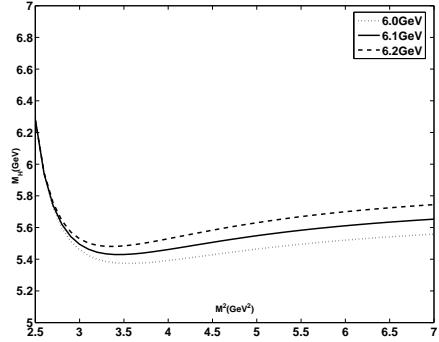}}\caption{
The mass of the $0^{+}$ tetraquark state with the scalar-scalar configuration as
a function of $M^2$ from sum rule (\ref{sum rule 1}). The continuum
thresholds are taken as $\sqrt{s_0}=6.0\sim6.2~\mbox{GeV}$. The
ranges of $M^{2}$ is $2.7\sim3.7~\mbox{GeV}^{2}$ for
$\sqrt{s_0}=6.0~\mbox{GeV}$, $2.7\sim3.8~\mbox{GeV}^{2}$ for
$\sqrt{s_0}=6.1~\mbox{GeV}$, and $2.7\sim3.9~\mbox{GeV}^{2}$ for
$\sqrt{s_0}=6.2~\mbox{GeV}$.}
\end{figure}

For the axial-axial configuration, the OPE contribution in sum rule
(\ref{sumrule1}) for $\sqrt{s_{0}}=6.1~\mbox{GeV}$
 is shown in Fig. 4 by comparing various contributions. Similarly, the two-quark condensate,
the four-quark condensate, and the mixed condensate
contributions could cancel each other out
to some extent and most
of the other condensates calculated are very small. Furthermore, the phenomenological contribution in sum rule
(\ref{sumrule1}) for $\sqrt{s_{0}}=6.1~\mbox{GeV}$ is displayed in Fig. 5.
The work windows for
the axial-axial case are taken as
$2.9\sim3.4~\mbox{GeV}^{2}$ for
$\sqrt{s_0}=6.0~\mbox{GeV}$, $2.9\sim3.5~\mbox{GeV}^{2}$ for
$\sqrt{s_0}=6.1~\mbox{GeV}$, and $2.9\sim3.6~\mbox{GeV}^{2}$ for
$\sqrt{s_0}=6.2~\mbox{GeV}$.
Its Borel curves are
shown in Fig.~6 and it is numerically evaluated as
$5.77\pm0.28~\mbox{GeV}$ in the work windows.
Considering the uncertainty from the variation of quark masses and
condensates, we get
$5.77\pm0.28^{+0.16}_{-0.05}~\mbox{GeV}$ (the
first error characterizes the uncertainty due to variation of $\sqrt{s_{0}}$
and $M^{2}$, and the second error is from the variation of
QCD parameters) for the axial-axial tetraquark state, or more concisely, $5.77^{+0.44}_{-0.33}~\mbox{GeV}$.

\begin{figure}
\centerline{\epsfysize=5.18truecm\epsfbox{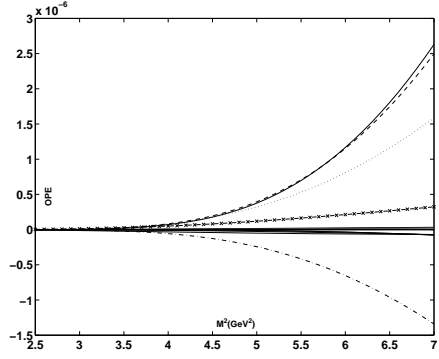}}
\caption{The various OPE contributions as a function of $M^2$ in sum rule
(\ref{sumrule1}) for $\sqrt{s_{0}}=6.1~\mbox{GeV}$ for the axial-axial case.
Four main contributions, i.e. the perturbative, the two-quark condensate,
the four-quark condensate, and the mixed condensate
are denoted by the single solid line,
the dashed line, the dotted line, and the dot-dashed line, respectively.
These main condensates could cancel each other out
to some extent, and
other condensates are very small.
All these factors mean that one can find perturbative dominance in the total
and  OPE convergence is still under control.}
\end{figure}

\begin{figure}
\centerline{\epsfysize=5.18truecm\epsfbox{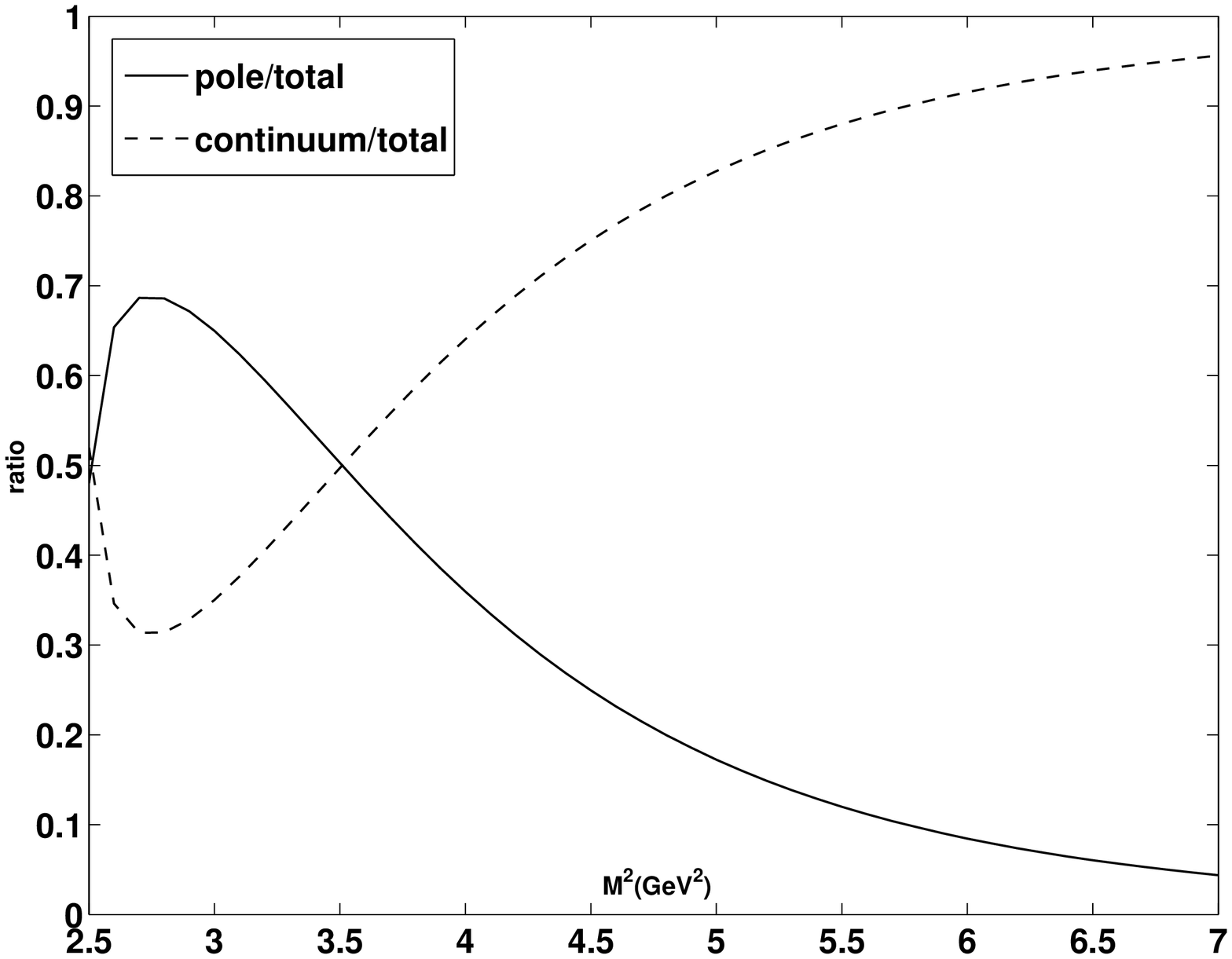}}
\caption{The phenomenological contribution in sum rule
(\ref{sumrule1}) for $\sqrt{s_{0}}=6.1~\mbox{GeV}$ for the axial-axial case.
The solid line is the relative pole contribution (the pole
contribution divided by the total, pole plus continuum contribution)
as a function of $M^2$ and the dashed line is the relative continuum
contribution.}
\end{figure}

\begin{figure}
\centerline{\epsfysize=5.18truecm
\epsfbox{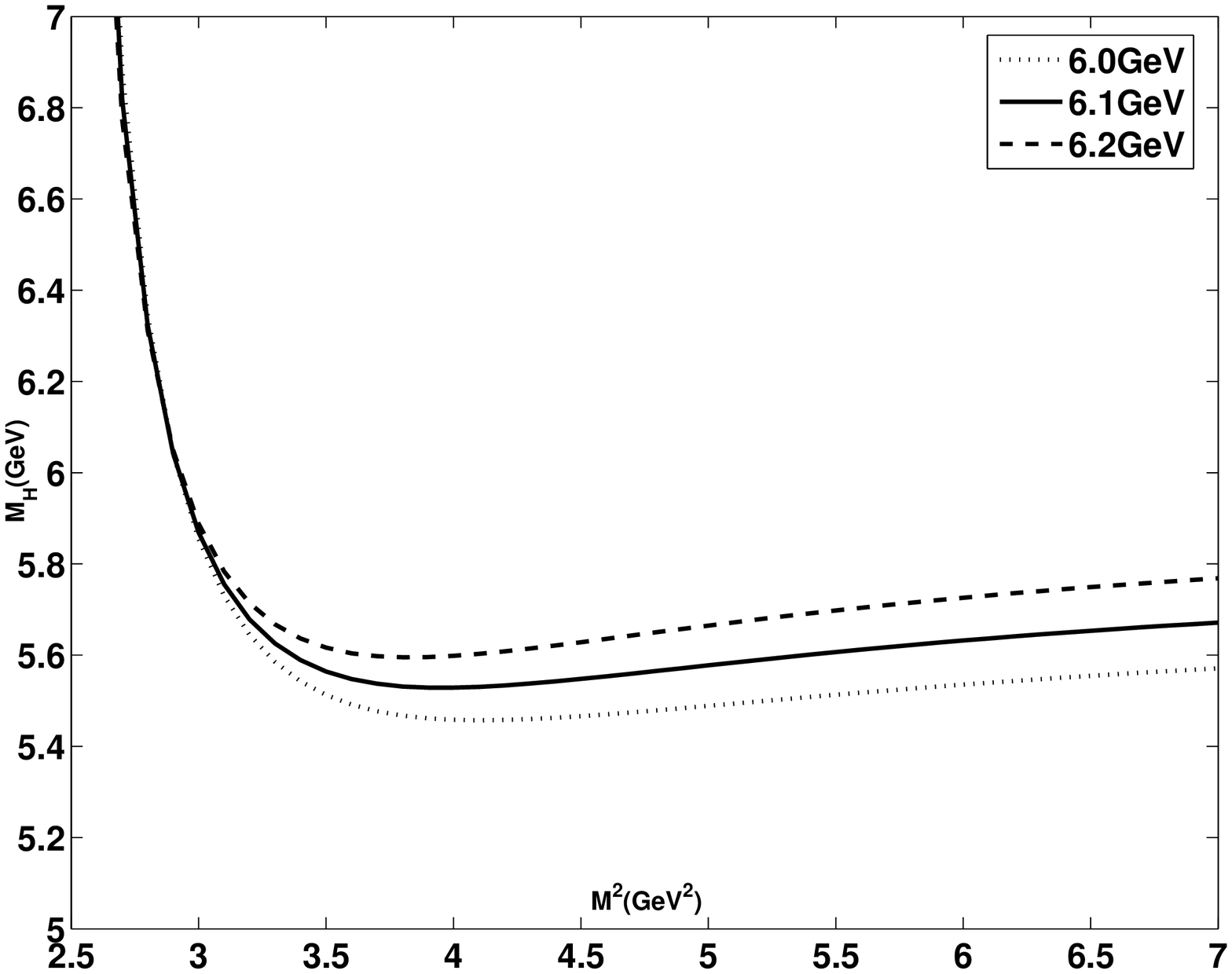}}\caption{
The mass of the $0^{+}$ tetraquark state
with the axial-axial configuration as
a function of $M^2$ from sum rule (\ref{sum rule 1}). The continuum
thresholds are taken as $\sqrt{s_0}=6.0\sim6.2~\mbox{GeV}$. The
ranges of $M^{2}$ are $2.9\sim3.4~\mbox{GeV}^{2}$ for
$\sqrt{s_0}=6.0~\mbox{GeV}$, $2.9\sim3.5~\mbox{GeV}^{2}$ for
$\sqrt{s_0}=6.1~\mbox{GeV}$, and $2.9\sim3.6~\mbox{GeV}^{2}$ for
$\sqrt{s_0}=6.2~\mbox{GeV}$.}
\end{figure}

For the pseudoscalar-pseudoscalar case, the OPE contribution in sum rule
(\ref{sumrule1}) for $\sqrt{s_{0}}=6.1~\mbox{GeV}$ is shown in Fig. 7.
There are also three main condensates, i.e. the two-quark condensate,
the four-quark condensate, and the mixed condensate on the OPE side.
They can certainly counteract each other
to some extent.
However, quite different from the  two cases discussed above,
there are two main condensates (the two-quark condensate
and the four-quark condensate) which
have a different sign from the perturbative term,
which means the signs of the perturbative part and the total OPE contribution are different.
The unsatisfactory OPE convergence means it is difficult to find some
reasonable work windows for this case.
It is inadvisable to keep on evaluating further numerical results.
Similarly, the OPE contribution in sum rule
(\ref{sumrule1}) for $\sqrt{s_{0}}=6.1~\mbox{GeV}$ for the vector-vector case is shown in Fig. 8.
It has the same problem as the pseudoscalar-pseudoscalar case,
and the most direct consequence
is that the Borel curves
are rather unstable.
Thus, it is also hard to find appropriate work windows to
extract credible hadronic information for the vector-vector case.

\begin{figure}
\centerline{\epsfysize=5.18truecm\epsfbox{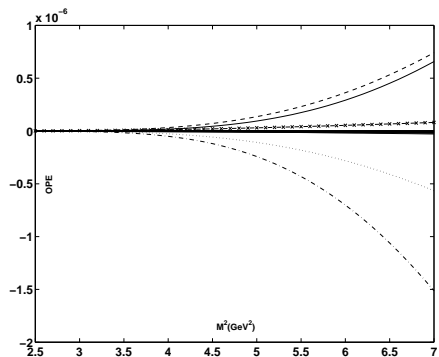}}
\caption{The various OPE contributions as a function of $M^2$ in sum rule
(\ref{sumrule1}) for $\sqrt{s_{0}}=6.1~\mbox{GeV}$ for the pseudoscalar-pseudoscalar case.
Four main contributions, i.e. the perturbative, the two-quark condensate,
the four-quark condensate, and the mixed condensate
are denoted by the single solid line, the dot-dashed line, the dotted line,
and the dashed line, respectively.
There two main condensates
have a different sign from the perturbative term,
which means the perturbative part and the total OPE contribution have different signs,
and the OPE convergence is not satisfactory for the pseudoscalar-pseudoscalar case.}
\end{figure}

\begin{figure}
\centerline{\epsfysize=5.18truecm\epsfbox{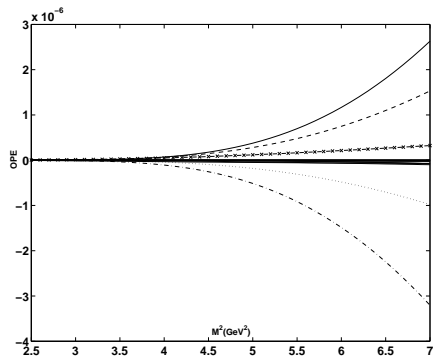}}
\caption{The various OPE contribution as a function of $M^2$ in sum rule
(\ref{sumrule1}) for $\sqrt{s_{0}}=6.1~\mbox{GeV}$ for the vector-vector case.
Four main contributions, i.e. the perturbative, the two-quark condensate,
the four-quark condensate, and the mixed condensate
are denoted by the single solid line, the dot-dashed line, the dotted line,
and the dashed line, respectively. There two main condensates
have a different sign from the perturbative term,
which means the perturbative part and the total OPE contribution have different signs,
and the OPE convergence is not satisfactory for the vector-vector case.}
\end{figure}

\section{Summary}\label{sec4}
Stimulated by the possibility of the recently observed
structure $X(5568)$ being an ideal candidate for exotic hadrons, we present an improved QCD sum rule study to
investigate whether $X(5568)$ could be a $0^{+}$ tetraquark state.
In deriving the sum rules, we have used four different interpolating currents, i.e.
the scalar-scalar,
the pseudoscalar-pseudoscalar, the axial-axial, and the vector-vector
diquark-antidiquark configurations.
Furthermore, in order to ensure the quality of QCD sum rule analysis, contributions of condensates
up to dimension $12$ are computed to test the OPE convergence.
We find that some condensates, such as the two-quark condensate,
the four-quark condensate, and the mixed condensate,
could play an important role on the OPE side.
The effect is not too bad for the scalar-scalar and the axial-axial cases, as their main condensates could cancel each other out
to some extent. Most
of the other condensates calculated are very small,
almost negligible,
which means that they cannot radically influence the character of OPE convergence.
All these factors mean that the OPE
convergence for the scalar-scalar and the axial-axial cases is still controllable.

Releasing the rigid OPE convergence criterion gives the following outcomes.
1) The final result for the scalar-scalar case is
$5.57^{+0.35}_{-0.23}~\mbox{GeV}$,
which coincides with the experimental data of $X(5568)$.
This result therefore supports
the tetraquark state explanation of $X(5568)$ with the scalar-scalar configuration.
2) For the axial-axial case, its eventual numerical value is
$5.77^{+0.44}_{-0.33}~\mbox{GeV}$, which is still consistent with the mass of $X(5568)$ in view of the uncertainty
although the central value is slightly higher.
Thus, one cannot arbitrarily exclude the possibility of
 $X(5568)$ being an axial-axial configuration tetraquark state.
3) For the pseudoscalar-pseudoscalar and the vector-vector cases,
their OPE convergence is so unsatisfactory that
one cannot find appropriate work windows to
get reliable hadronic information.
In future, it is expected that more precise information on the nature of the $X(5568)$
will be revealed by the further contributions of both experimental observations
and theoretical studies.

\begin{acknowledgments}
This work was supported by the National
Natural Science Foundation of China under Contract
Nos. 11475258, 11105223, and 11675263, and the project in
NUDT for excellent youth talents.
\end{acknowledgments}

\begin{appendix}
\section{}

The
spectral density $\rho_{i}(s)$ is
\begin{eqnarray}
\rho_{i}(s)=\rho_{i}^{\mbox{pert}}(s)+\rho_{i}^{\langle\bar{q}q\rangle}(s)+\rho_{i}^{\langle
g^{2}G^{2}\rangle}(s)+\rho_{i}^{\langle
g\bar{q}\sigma\cdot G q\rangle}(s)+\rho_{i}^{\langle\bar{q}q\rangle^{2}}(s)+\rho_{i}^{\langle
g^{3}G^{3}\rangle}(s)+\rho_{i}^{\langle\bar{q}q\rangle\langle
g^{2}G^{2}\rangle}(s),\nonumber
\end{eqnarray}
with $i=I$, $II$, $III$, and $IV$.
Concretely,
\begin{eqnarray}
\rho_{I}^{\mbox{pert}}(s)&=&\frac{1}{3\cdot2^{10}\pi^{6}}\int_{\Lambda}^{1}d\alpha\bigg(\frac{1-\alpha}{\alpha}\bigg)^{3}(\alpha s-m_{Q}^{2})^{4},\nonumber\\
\rho_{II}^{\mbox{pert}}(s)&=&\frac{1}{3\cdot2^{10}\pi^{6}}\int_{\Lambda}^{1}d\alpha\bigg(\frac{1-\alpha}{\alpha}\bigg)^{3}(\alpha s-m_{Q}^{2})^{4},\nonumber\\
\rho_{III}^{\mbox{pert}}(s)&=&\frac{1}{3\cdot2^{8}\pi^{6}}\int_{\Lambda}^{1}d\alpha\bigg(\frac{1-\alpha}{\alpha}\bigg)^{3}(\alpha s-m_{Q}^{2})^{4},\nonumber\\
\rho_{IV}^{\mbox{pert}}(s)&=&\frac{1}{3\cdot2^{8}\pi^{6}}\int_{\Lambda}^{1}d\alpha\bigg(\frac{1-\alpha}{\alpha}\bigg)^{3}(\alpha s-m_{Q}^{2})^{4},\nonumber\\
\rho_{I}^{\langle\bar{q}q\rangle}(s)&=&\frac{1}{2^{6}\pi^{4}}\int_{\Lambda}^{1}d\alpha\bigg(m_{s}\langle\bar{s}s\rangle-2m_{s}\langle\bar{q}q\rangle -m_{Q}\langle\bar{q}q\rangle\frac{1-\alpha}{\alpha}\bigg)\frac{1-\alpha}{\alpha}(\alpha s-m_{Q}^{2})^{2},\nonumber\\
\rho_{II}^{\langle\bar{q}q\rangle}(s)&=&\frac{1}{2^{6}\pi^{4}}\int_{\Lambda}^{1}d\alpha\bigg(m_{s}\langle\bar{s}s\rangle+2m_{s}\langle\bar{q}q\rangle +m_{Q}\langle\bar{q}q\rangle\frac{1-\alpha}{\alpha}\bigg)\frac{1-\alpha}{\alpha}(\alpha s-m_{Q}^{2})^{2},\nonumber\\
\rho_{III}^{\langle\bar{q}q\rangle}(s)&=&\frac{1}{2^{4}\pi^{4}}\int_{\Lambda}^{1}d\alpha\bigg(m_{s}\langle\bar{s}s\rangle-m_{s}\langle\bar{q}q\rangle -\frac{1}{2}m_{Q}\langle\bar{q}q\rangle\frac{1-\alpha}{\alpha}\bigg)\frac{1-\alpha}{\alpha}(\alpha s-m_{Q}^{2})^{2},\nonumber\\
\rho_{IV}^{\langle\bar{q}q\rangle}(s)&=&\frac{1}{2^{4}\pi^{4}}\int_{\Lambda}^{1}d\alpha\bigg(m_{s}\langle\bar{s}s\rangle+m_{s}\langle\bar{q}q\rangle +\frac{1}{2}m_{Q}\langle\bar{q}q\rangle\frac{1-\alpha}{\alpha}\bigg)\frac{1-\alpha}{\alpha}(\alpha s-m_{Q}^{2})^{2},\nonumber\\
\rho_{I}^{\langle g^{2}G^{2}\rangle}(s)&=&-\frac{m_{Q}^{2}\langle
g^{2}G^{2}\rangle}{3^{2}\cdot2^{10}\pi^{6}}\int_{\Lambda}^{1}d\alpha\bigg(\frac{1-\alpha}{\alpha}\bigg)^{3}(\alpha s-m_{Q}^{2}),\nonumber\\
\rho_{II}^{\langle g^{2}G^{2}\rangle}(s)&=&-\frac{m_{Q}^{2}\langle
g^{2}G^{2}\rangle}{3^{2}\cdot2^{10}\pi^{6}}\int_{\Lambda}^{1}d\alpha\bigg(\frac{1-\alpha}{\alpha}\bigg)^{3}(\alpha s-m_{Q}^{2}),\nonumber\\
\rho_{III}^{\langle g^{2}G^{2}\rangle}(s)&=&-\frac{m_{Q}^{2}\langle
g^{2}G^{2}\rangle}{3^{2}\cdot2^{8}\pi^{6}}\int_{\Lambda}^{1}d\alpha\bigg(\frac{1-\alpha}{\alpha}\bigg)^{3}(\alpha s-m_{Q}^{2}),\nonumber\\
\rho_{IV}^{\langle g^{2}G^{2}\rangle}(s)&=&-\frac{m_{Q}^{2}\langle
g^{2}G^{2}\rangle}{3^{2}\cdot2^{8}\pi^{6}}\int_{\Lambda}^{1}d\alpha\bigg(\frac{1-\alpha}{\alpha}\bigg)^{3}(\alpha s-m_{Q}^{2}),\nonumber\\
\rho_{I}^{\langle g\bar{q}\sigma\cdot G q\rangle}(s)&=&-\frac{1}{3\cdot2^{6}\pi^{4}}\int_{\Lambda}^{1}d\alpha\bigg(m_{s}\langle
g\bar{s}\sigma\cdot G
s\rangle-3m_{s}\langle
g\bar{q}\sigma\cdot G
q\rangle-3m_{Q}\langle
g\bar{q}\sigma\cdot G
q\rangle\frac{1-\alpha}{\alpha}\bigg)(\alpha s-m_{Q}^{2}),\nonumber\\
\rho_{II}^{\langle g\bar{q}\sigma\cdot G q\rangle}(s)&=&-\frac{1}{3\cdot2^{6}\pi^{4}}\int_{\Lambda}^{1}d\alpha\bigg(m_{s}\langle
g\bar{s}\sigma\cdot G
s\rangle+3m_{s}\langle
g\bar{q}\sigma\cdot G
q\rangle+3m_{Q}\langle
g\bar{q}\sigma\cdot G
q\rangle\frac{1-\alpha}{\alpha}\bigg)(\alpha s-m_{Q}^{2}),\nonumber\\
\rho_{III}^{\langle g\bar{q}\sigma\cdot G q\rangle}(s)&=&-\frac{1}{3\cdot2^{4}\pi^{4}}\int_{\Lambda}^{1}d\alpha\bigg(m_{s}\langle
g\bar{s}\sigma\cdot G
s\rangle-\frac{3}{2}m_{s}\langle
g\bar{q}\sigma\cdot G
q\rangle-\frac{3}{2}m_{Q}\langle
g\bar{q}\sigma\cdot G
q\rangle\frac{1-\alpha}{\alpha}\bigg)(\alpha s-m_{Q}^{2}),\nonumber\\
\rho_{IV}^{\langle g\bar{q}\sigma\cdot G q\rangle}(s)&=&-\frac{1}{3\cdot2^{4}\pi^{4}}\int_{\Lambda}^{1}d\alpha\bigg(m_{s}\langle
g\bar{s}\sigma\cdot G
s\rangle+\frac{3}{2}m_{s}\langle
g\bar{q}\sigma\cdot G
q\rangle+\frac{3}{2}m_{Q}\langle
g\bar{q}\sigma\cdot G
q\rangle\frac{1-\alpha}{\alpha}\bigg)(\alpha s-m_{Q}^{2}),\nonumber\\
\rho_{I}^{\langle\bar{q}q\rangle^{2}}(s)&=&\frac{1}{3\cdot2^{3}\pi^{2}}\int_{\Lambda}^{1}d\alpha\langle\bar{q}q\rangle\Big[2\langle\bar{s}s\rangle(\alpha s-m_{Q}^{2})+2m_{Q}m_{s}\langle\bar{q}q\rangle-m_{Q}m_{s}\langle\bar{s}s\rangle\Big],\nonumber\\
\rho_{II}^{\langle\bar{q}q\rangle^{2}}(s)&=&\frac{1}{3\cdot2^{3}\pi^{2}}\int_{\Lambda}^{1}d\alpha\langle\bar{q}q\rangle\Big[-2\langle\bar{s}s\rangle(\alpha s-m_{Q}^{2})+2m_{Q}m_{s}\langle\bar{q}q\rangle+m_{Q}m_{s}\langle\bar{s}s\rangle\Big],\nonumber\\
\rho_{III}^{\langle\bar{q}q\rangle^{2}}(s)&=&\frac{1}{3\cdot2^{2}\pi^{2}}\int_{\Lambda}^{1}d\alpha\langle\bar{q}q\rangle\Big[2\langle\bar{s}s\rangle(\alpha s-m_{Q}^{2})+4m_{Q}m_{s}\langle\bar{q}q\rangle-m_{Q}m_{s}\langle\bar{s}s\rangle\Big],\nonumber\\
\rho_{IV}^{\langle\bar{q}q\rangle^{2}}(s)&=&\frac{1}{3\cdot2^{2}\pi^{2}}\int_{\Lambda}^{1}d\alpha\langle\bar{q}q\rangle\Big[-2\langle\bar{s}s\rangle(\alpha s-m_{Q}^{2})+4m_{Q}m_{s}\langle\bar{q}q\rangle+m_{Q}m_{s}\langle\bar{s}s\rangle\Big],\nonumber\\
\rho_{I}^{\langle g^{3}G^{3}\rangle}(s)&=&-\frac{\langle
g^{3}G^{3}\rangle}{3^{2}\cdot2^{12}\pi^{6}}\int_{\Lambda}^{1}d\alpha\bigg(\frac{1-\alpha}{\alpha}\bigg)^{3}(\alpha s-3m_{Q}^{2}),\nonumber\\
\rho_{II}^{\langle g^{3}G^{3}\rangle}(s)&=&-\frac{\langle
g^{3}G^{3}\rangle}{3^{2}\cdot2^{12}\pi^{6}}\int_{\Lambda}^{1}d\alpha\bigg(\frac{1-\alpha}{\alpha}\bigg)^{3}(\alpha s-3m_{Q}^{2}),\nonumber\\
\rho_{III}^{\langle g^{3}G^{3}\rangle}(s)&=&-\frac{\langle
g^{3}G^{3}\rangle}{3^{2}\cdot2^{10}\pi^{6}}\int_{\Lambda}^{1}d\alpha\bigg(\frac{1-\alpha}{\alpha}\bigg)^{3}(\alpha s-3m_{Q}^{2}),\nonumber\\
\rho_{IV}^{\langle g^{3}G^{3}\rangle}(s)&=&-\frac{\langle
g^{3}G^{3}\rangle}{3^{2}\cdot2^{10}\pi^{6}}\int_{\Lambda}^{1}d\alpha\bigg(\frac{1-\alpha}{\alpha}\bigg)^{3}(\alpha s-3m_{Q}^{2}),\nonumber\\
\rho_{I}^{\langle\bar{q}q\rangle\langle
g^{2}G^{2}\rangle}(s)&=&-\frac{m_{Q}\langle\bar{q}q\rangle\langle
g^{2}G^{2}\rangle}{3^{2}\cdot2^{8}\pi^{4}}\int_{\Lambda}^{1}d\alpha\bigg[1+3\bigg(\frac{1-\alpha}{\alpha}\bigg)^{2}\bigg],\nonumber\\
\rho_{II}^{\langle\bar{q}q\rangle\langle
g^{2}G^{2}\rangle}(s)&=&\frac{m_{Q}\langle\bar{q}q\rangle\langle
g^{2}G^{2}\rangle}{3^{2}\cdot2^{8}\pi^{4}}\int_{\Lambda}^{1}d\alpha\bigg[1+3\bigg(\frac{1-\alpha}{\alpha}\bigg)^{2}\bigg],\nonumber\\
\rho_{III}^{\langle\bar{q}q\rangle\langle
g^{2}G^{2}\rangle}(s)&=&-\frac{m_{Q}\langle\bar{q}q\rangle\langle
g^{2}G^{2}\rangle}{3^{2}\cdot2^{7}\pi^{4}}\int_{\Lambda}^{1}d\alpha\bigg[1+3\bigg(\frac{1-\alpha}{\alpha}\bigg)^{2}\bigg],\nonumber
\end{eqnarray}
and
\begin{eqnarray}
\rho_{IV}^{\langle\bar{q}q\rangle\langle
g^{2}G^{2}\rangle}(s)&=&\frac{m_{Q}\langle\bar{q}q\rangle\langle
g^{2}G^{2}\rangle}{3^{2}\cdot2^{7}\pi^{4}}\int_{\Lambda}^{1}d\alpha\bigg[1+3\bigg(\frac{1-\alpha}{\alpha}\bigg)^{2}\bigg],\nonumber
\end{eqnarray}
with $\Lambda=m_{Q}^2/s$.

The $\hat{B}\Pi_{i}^{\mbox{cond}}$ term reads
\begin{eqnarray}
\hat{B}\Pi_{i}^{\mbox{cond}}&=&\hat{B}\Pi_{i}^{\langle\bar{q}q\rangle^{3}}+\hat{B}\Pi_{i}^{\langle\bar{q}q\rangle\langle g\bar{q}\sigma\cdot G q\rangle}+\hat{B}\Pi_{i}^{\langle g\bar{q}\sigma\cdot G q\rangle^{2}}+\hat{B}\Pi_{i}^{\langle\bar{q}q\rangle\langle
g^{2}G^{2}\rangle}+\hat{B}\Pi_{i}^{\langle\bar{q}q\rangle\langle
g^{3}G^{3}\rangle}+\hat{B}\Pi_{i}^{\langle g^{2}G^{2}\rangle\langle g\bar{q}\sigma\cdot G q\rangle}\nonumber\\
&+&\hat{B}\Pi_{i}^{\langle\bar{q}q\rangle^{2}\langle g\bar{q}\sigma\cdot G q\rangle}+\hat{B}\Pi_{i}^{\langle\bar{q}q\rangle\langle g^{2}G^{2}\rangle^{2}}+\hat{B}\Pi_{i}^{\langle\bar{q}q\rangle^{2}\langle g^{2}G^{2}\rangle}+\hat{B}\Pi_{i}^{\langle\bar{q}q\rangle^{2}\langle g^{3}G^{3}\rangle}+\hat{B}\Pi_{i}^{\langle\bar{q}q\rangle\langle g^{2}G^{2}\rangle\langle g\bar{q}\sigma\cdot G q\rangle}+\hat{B}\Pi_{i}^{\langle g^{3}G^{3}\rangle\langle g\bar{q}\sigma\cdot G q\rangle},\nonumber
\end{eqnarray}
with $i=I$, $II$, $III$, and $IV$.
One by one,
\begin{eqnarray}
\hat{B}\Pi_{I}^{\langle\bar{q}q\rangle^{3}}&=&-\frac{m_{Q}}{3^{2}}\langle\bar{q}q\rangle^{2}\langle\bar{s}s\rangle e^{-\frac{m_{Q}^{2}}{M^{2}}},\nonumber\\
\hat{B}\Pi_{II}^{\langle\bar{q}q\rangle^{3}}&=&-\frac{m_{Q}}{3^{2}}\langle\bar{q}q\rangle^{2}\langle\bar{s}s\rangle e^{-\frac{m_{Q}^{2}}{M^{2}}},\nonumber\\
\hat{B}\Pi_{III}^{\langle\bar{q}q\rangle^{3}}&=&-\frac{2^{2}m_{Q}}{3^{2}}\langle\bar{q}q\rangle^{2}\langle\bar{s}s\rangle e^{-\frac{m_{Q}^{2}}{M^{2}}},\nonumber\\
\hat{B}\Pi_{IV}^{\langle\bar{q}q\rangle^{3}}&=&-\frac{2^{2}m_{Q}}{3^{2}}\langle\bar{q}q\rangle^{2}\langle\bar{s}s\rangle e^{-\frac{m_{Q}^{2}}{M^{2}}},\nonumber\\
\hat{B}\Pi_{I}^{\langle\bar{q}q\rangle\langle g\bar{q}\sigma\cdot G q\rangle}&=&-\frac{m_{Q}m_{s}}{3^{2}\cdot2^{5}\pi^{2}}\bigg(-2\langle\bar{q}q\rangle\langle g\bar{s}\sigma\cdot G s\rangle+12\langle\bar{q}q\rangle\langle g\bar{q}\sigma\cdot G q\rangle-3\langle\bar{s}s\rangle\langle g\bar{q}\sigma\cdot G q\rangle\bigg)e^{-\frac{m_{Q}^{2}}{M^{2}}},\nonumber\\
\hat{B}\Pi_{II}^{\langle\bar{q}q\rangle\langle g\bar{q}\sigma\cdot G q\rangle}&=&-\frac{m_{Q}m_{s}}{3^{2}\cdot2^{5}\pi^{2}}\bigg(2\langle\bar{q}q\rangle\langle g\bar{s}\sigma\cdot G s\rangle+12\langle\bar{q}q\rangle\langle g\bar{q}\sigma\cdot G q\rangle+3\langle\bar{s}s\rangle\langle g\bar{q}\sigma\cdot G q\rangle\bigg)e^{-\frac{m_{Q}^{2}}{M^{2}}},\nonumber\\
\hat{B}\Pi_{III}^{\langle\bar{q}q\rangle\langle g\bar{q}\sigma\cdot G q\rangle}&=&-\frac{m_{Q}m_{s}}{3^{2}\cdot2^{4}\pi^{2}}\bigg(-2\langle\bar{q}q\rangle\langle g\bar{s}\sigma\cdot G s\rangle+24\langle\bar{q}q\rangle\langle g\bar{q}\sigma\cdot G q\rangle-3\langle\bar{s}s\rangle\langle g\bar{q}\sigma\cdot G q\rangle\bigg)e^{-\frac{m_{Q}^{2}}{M^{2}}},\nonumber\\
\hat{B}\Pi_{IV}^{\langle\bar{q}q\rangle\langle g\bar{q}\sigma\cdot G q\rangle}&=&-\frac{m_{Q}m_{s}}{3^{2}\cdot2^{4}\pi^{2}}\bigg(2\langle\bar{q}q\rangle\langle g\bar{s}\sigma\cdot G s\rangle+24\langle\bar{q}q\rangle\langle g\bar{q}\sigma\cdot G q\rangle+3\langle\bar{s}s\rangle\langle g\bar{q}\sigma\cdot G q\rangle\bigg)e^{-\frac{m_{Q}^{2}}{M^{2}}},\nonumber\\
\hat{B}\Pi_{I}^{\langle g\bar{q}\sigma\cdot G q\rangle^{2}}&=&\frac{\langle g\bar{q}\sigma\cdot G q\rangle}{3^{2}\cdot2^{6}\pi^{2}}\bigg[3\langle g\bar{s}\sigma\cdot G s\rangle\bigg(1+\frac{m_{Q}^{2}}{M^{2}}\bigg)
-m_{s}m_{Q}^{3}\Big(\langle g\bar{s}\sigma\cdot G s\rangle-3\langle g\bar{q}\sigma\cdot G q\rangle\Big)\frac{1}{(M^{2})^{2}}\bigg]e^{-\frac{m_{Q}^{2}}{M^{2}}},\nonumber\\
\hat{B}\Pi_{II}^{\langle g\bar{q}\sigma\cdot G q\rangle^{2}}&=&\frac{\langle g\bar{q}\sigma\cdot G q\rangle}{3^{2}\cdot2^{6}\pi^{2}}\bigg[-3\langle g\bar{s}\sigma\cdot G s\rangle\bigg(1+\frac{m_{Q}^{2}}{M^{2}}\bigg)
+m_{s}m_{Q}^{3}\Big(\langle g\bar{s}\sigma\cdot G s\rangle+3\langle g\bar{q}\sigma\cdot G q\rangle\Big)\frac{1}{(M^{2})^{2}}\bigg]e^{-\frac{m_{Q}^{2}}{M^{2}}},\nonumber\\
\hat{B}\Pi_{III}^{\langle g\bar{q}\sigma\cdot G q\rangle^{2}}&=&\frac{\langle g\bar{q}\sigma\cdot G q\rangle}{3^{2}\cdot2^{5}\pi^{2}}\bigg[3\langle g\bar{s}\sigma\cdot G s\rangle\bigg(1+\frac{m_{Q}^{2}}{M^{2}}\bigg)
-m_{s}m_{Q}^{3}\Big(\langle g\bar{s}\sigma\cdot G s\rangle-6\langle g\bar{q}\sigma\cdot G q\rangle\Big)\frac{1}{(M^{2})^{2}}\bigg]e^{-\frac{m_{Q}^{2}}{M^{2}}},\nonumber\\
\hat{B}\Pi_{IV}^{\langle g\bar{q}\sigma\cdot G q\rangle^{2}}&=&\frac{\langle g\bar{q}\sigma\cdot G q\rangle}{3^{2}\cdot2^{5}\pi^{2}}\bigg[-3\langle g\bar{s}\sigma\cdot G s\rangle\bigg(1+\frac{m_{Q}^{2}}{M^{2}}\bigg)
+m_{s}m_{Q}^{3}\Big(\langle g\bar{s}\sigma\cdot G s\rangle+6\langle g\bar{q}\sigma\cdot G q\rangle\Big)\frac{1}{(M^{2})^{2}}\bigg]e^{-\frac{m_{Q}^{2}}{M^{2}}},\nonumber\\
\hat{B}\Pi_{I}^{\langle\bar{q}q\rangle\langle
g^{2}G^{2}\rangle}&=&\frac{m_{Q}^{2}\langle
g^{2}G^{2}\rangle}{3^{2}\cdot2^{8}\pi^{4}}\int_{0}^{1}d\alpha\bigg[m_{Q}\langle\bar{q}q\rangle\frac{1-\alpha}{\alpha}-m_{s}\Big(\langle\bar{s}s\rangle-2\langle\bar{q}q\rangle\Big)\bigg]
\frac{1-\alpha}{\alpha^{2}}e^{-\frac{m_{Q}^{2}}{\alpha M^{2}}},\nonumber\\
\hat{B}\Pi_{II}^{\langle\bar{q}q\rangle\langle
g^{2}G^{2}\rangle}&=&-\frac{m_{Q}^{2}\langle
g^{2}G^{2}\rangle}{3^{2}\cdot2^{8}\pi^{4}}\int_{0}^{1}d\alpha\bigg[m_{Q}\langle\bar{q}q\rangle\frac{1-\alpha}{\alpha}+m_{s}\Big(\langle\bar{s}s\rangle+2\langle\bar{q}q\rangle\Big)\bigg]
\frac{1-\alpha}{\alpha^{2}}e^{-\frac{m_{Q}^{2}}{\alpha M^{2}}},\nonumber\\
\hat{B}\Pi_{III}^{\langle\bar{q}q\rangle\langle
g^{2}G^{2}\rangle}&=&\frac{m_{Q}^{2}\langle
g^{2}G^{2}\rangle}{3^{2}\cdot2^{6}\pi^{4}}\int_{0}^{1}d\alpha\bigg[\frac{1}{2}m_{Q}\langle\bar{q}q\rangle\frac{1-\alpha}{\alpha}-m_{s}\Big(\langle\bar{s}s\rangle-\langle\bar{q}q\rangle\Big)\bigg]
\frac{1-\alpha}{\alpha^{2}}e^{-\frac{m_{Q}^{2}}{\alpha M^{2}}},\nonumber\\
\hat{B}\Pi_{IV}^{\langle\bar{q}q\rangle\langle
g^{2}G^{2}\rangle}&=&-\frac{m_{Q}^{2}\langle
g^{2}G^{2}\rangle}{3^{2}\cdot2^{6}\pi^{4}}\int_{0}^{1}d\alpha\bigg[\frac{1}{2}m_{Q}\langle\bar{q}q\rangle\frac{1-\alpha}{\alpha}+m_{s}\Big(\langle\bar{s}s\rangle+\langle\bar{q}q\rangle\Big)\bigg]
\frac{1-\alpha}{\alpha^{2}}e^{-\frac{m_{Q}^{2}}{\alpha M^{2}}},\nonumber\\
\hat{B}\Pi_{I}^{\langle\bar{q}q\rangle\langle
g^{3}G^{3}\rangle}&=&-\frac{\langle
g^{3}G^{3}\rangle}{3^{2}\cdot2^{10}\pi^{4}}\int_{0}^{1}d\alpha\bigg[ m_{s}\Big(\langle\bar{s}s\rangle-2\langle\bar{q}q\rangle\Big)\bigg(1-2\frac{m_{Q}^{2}}{\alpha M^{2}}\bigg)\nonumber\\
&-&2m_{Q}\langle\bar{q}q\rangle\bigg(3-\frac{m_{Q}^{2}}{\alpha M^{2}}\bigg)\frac{1-\alpha}{\alpha}\bigg]\frac{1-\alpha}{\alpha^{2}}e^{-\frac{m_{Q}^{2}}{\alpha M^{2}}},\nonumber\\
\hat{B}\Pi_{II}^{\langle\bar{q}q\rangle\langle
g^{3}G^{3}\rangle}&=&-\frac{\langle
g^{3}G^{3}\rangle}{3^{2}\cdot2^{10}\pi^{4}}\int_{0}^{1}d\alpha\bigg[ m_{s}\Big(\langle\bar{s}s\rangle+2\langle\bar{q}q\rangle\Big)\bigg(1-2\frac{m_{Q}^{2}}{\alpha M^{2}}\bigg)\nonumber\\
&+&2m_{Q}\langle\bar{q}q\rangle\bigg(3-\frac{m_{Q}^{2}}{\alpha M^{2}}\bigg)\frac{1-\alpha}{\alpha}\bigg]\frac{1-\alpha}{\alpha^{2}}e^{-\frac{m_{Q}^{2}}{\alpha M^{2}}},\nonumber\\
\hat{B}\Pi_{III}^{\langle\bar{q}q\rangle\langle
g^{3}G^{3}\rangle}&=&-\frac{\langle
g^{3}G^{3}\rangle}{3^{2}\cdot2^{8}\pi^{4}}\int_{0}^{1}d\alpha\bigg[ m_{s}\Big(\langle\bar{s}s\rangle-\langle\bar{q}q\rangle\Big)\bigg(1-2\frac{m_{Q}^{2}}{\alpha M^{2}}\bigg)\nonumber\\
&-&m_{Q}\langle\bar{q}q\rangle\bigg(3-\frac{m_{Q}^{2}}{\alpha M^{2}}\bigg)\frac{1-\alpha}{\alpha}\bigg]\frac{1-\alpha}{\alpha^{2}}e^{-\frac{m_{Q}^{2}}{\alpha M^{2}}},\nonumber\\
\hat{B}\Pi_{IV}^{\langle\bar{q}q\rangle\langle
g^{3}G^{3}\rangle}&=&-\frac{\langle
g^{3}G^{3}\rangle}{3^{2}\cdot2^{8}\pi^{4}}\int_{0}^{1}d\alpha\bigg[ m_{s}\Big(\langle\bar{s}s\rangle+\langle\bar{q}q\rangle\Big)\bigg(1-2\frac{m_{Q}^{2}}{\alpha M^{2}}\bigg)\nonumber\\
&+&m_{Q}\langle\bar{q}q\rangle\bigg(3-\frac{m_{Q}^{2}}{\alpha M^{2}}\bigg)\frac{1-\alpha}{\alpha}\bigg]\frac{1-\alpha}{\alpha^{2}}e^{-\frac{m_{Q}^{2}}{\alpha M^{2}}},\nonumber\\
\hat{B}\Pi_{I}^{\langle g^{2}G^{2}\rangle\langle g\bar{q}\sigma\cdot G q\rangle}&=&\frac{m_{Q}\langle g^{2}G^{2}\rangle}{3^{3}\cdot2^{9}\pi^{4}}\int_{0}^{1}\frac{d\alpha}{\alpha^{2}}\bigg[\frac{m_{Q}m_{s}\Big(\langle g\bar{s}\sigma\cdot G s\rangle-3\langle g\bar{q}\sigma\cdot G q\rangle\Big)}{M^{2}}\nonumber\\
&+&3\langle g\bar{q}\sigma\cdot G q\rangle(1-\alpha)\bigg(3-\frac{m_{Q}^{2}}{\alpha M^{2}}\bigg)\bigg]e^{-\frac{m_{Q}^{2}}{\alpha M^{2}}},\nonumber\\
\hat{B}\Pi_{II}^{\langle g^{2}G^{2}\rangle\langle g\bar{q}\sigma\cdot G q\rangle}&=&\frac{m_{Q}\langle g^{2}G^{2}\rangle}{3^{3}\cdot2^{9}\pi^{4}}\int_{0}^{1}\frac{d\alpha}{\alpha^{2}}\bigg[\frac{m_{Q}m_{s}\Big(\langle g\bar{s}\sigma\cdot G s\rangle+3\langle g\bar{q}\sigma\cdot G q\rangle\Big)}{M^{2}}\nonumber\\
&-&3\langle g\bar{q}\sigma\cdot G q\rangle(1-\alpha)\bigg(3-\frac{m_{Q}^{2}}{\alpha M^{2}}\bigg)\bigg]e^{-\frac{m_{Q}^{2}}{\alpha M^{2}}},\nonumber\\
\hat{B}\Pi_{III}^{\langle g^{2}G^{2}\rangle\langle g\bar{q}\sigma\cdot G q\rangle}&=&\frac{m_{Q}\langle g^{2}G^{2}\rangle}{3^{3}\cdot2^{7}\pi^{4}}\int_{0}^{1}\frac{d\alpha}{\alpha^{2}}\bigg[\frac{m_{Q}m_{s}\Big(\langle g\bar{s}\sigma\cdot G s\rangle-\frac{3}{2}\langle g\bar{q}\sigma\cdot G q\rangle\Big)}{M^{2}}\nonumber\\
&+&\frac{3}{2}\langle g\bar{q}\sigma\cdot G q\rangle(1-\alpha)\bigg(3-\frac{m_{Q}^{2}}{\alpha M^{2}}\bigg)\bigg]e^{-\frac{m_{Q}^{2}}{\alpha M^{2}}},\nonumber\\
\hat{B}\Pi_{IV}^{\langle g^{2}G^{2}\rangle\langle g\bar{q}\sigma\cdot G q\rangle}&=&\frac{m_{Q}\langle g^{2}G^{2}\rangle}{3^{3}\cdot2^{7}\pi^{4}}\int_{0}^{1}\frac{d\alpha}{\alpha^{2}}\bigg[\frac{m_{Q}m_{s}\Big(\langle g\bar{s}\sigma\cdot G s\rangle+\frac{3}{2}\langle g\bar{q}\sigma\cdot G q\rangle\Big)}{M^{2}}\nonumber\\
&-&\frac{3}{2}\langle g\bar{q}\sigma\cdot G q\rangle(1-\alpha)\bigg(3-\frac{m_{Q}^{2}}{\alpha M^{2}}\bigg)\bigg]e^{-\frac{m_{Q}^{2}}{\alpha M^{2}}},\nonumber\\
\hat{B}\Pi_{I}^{\langle\bar{q}q\rangle^{2}\langle g\bar{q}\sigma\cdot G q\rangle}&=&\frac{m_{Q}^{3}\langle\bar{q}q\rangle\Big(\langle\bar{q}q\rangle\langle g\bar{s}\sigma\cdot G s\rangle+2\langle\bar{s}s\rangle\langle g\bar{q}\sigma\cdot G q\rangle\Big)}{3^{2}\cdot2^{2}(M^{2})^{2}}e^{-\frac{m_{Q}^{2}}{M^{2}}},\nonumber\\
\hat{B}\Pi_{II}^{\langle\bar{q}q\rangle^{2}\langle g\bar{q}\sigma\cdot G q\rangle}&=&\frac{m_{Q}^{3}\langle\bar{q}q\rangle\Big(\langle\bar{q}q\rangle\langle g\bar{s}\sigma\cdot G s\rangle+2\langle\bar{s}s\rangle\langle g\bar{q}\sigma\cdot G q\rangle\Big)}{3^{2}\cdot2^{2}(M^{2})^{2}}e^{-\frac{m_{Q}^{2}}{M^{2}}},\nonumber\\
\hat{B}\Pi_{III}^{\langle\bar{q}q\rangle^{2}\langle g\bar{q}\sigma\cdot G q\rangle}&=&\frac{m_{Q}^{3}\langle\bar{q}q\rangle\Big(\langle\bar{q}q\rangle\langle g\bar{s}\sigma\cdot G s\rangle+2\langle\bar{s}s\rangle\langle g\bar{q}\sigma\cdot G q\rangle\Big)}{3^{2}(M^{2})^{2}}e^{-\frac{m_{Q}^{2}}{M^{2}}},\nonumber\\
\hat{B}\Pi_{IV}^{\langle\bar{q}q\rangle^{2}\langle g\bar{q}\sigma\cdot G q\rangle}&=&\frac{m_{Q}^{3}\langle\bar{q}q\rangle\Big(\langle\bar{q}q\rangle\langle g\bar{s}\sigma\cdot G s\rangle+2\langle\bar{s}s\rangle\langle g\bar{q}\sigma\cdot G q\rangle\Big)}{3^{2}(M^{2})^{2}}e^{-\frac{m_{Q}^{2}}{M^{2}}},\nonumber\\
\hat{B}\Pi_{I}^{\langle\bar{q}q\rangle\langle g^{2}G^{2}\rangle^{2}}&=&-\frac{m_{Q}\langle\bar{q}q\rangle\langle g^{2}G^{2}\rangle^{2}}{3^{4}\cdot2^{11}\pi^{4}}\int_{0}^{1}d\alpha
\bigg(3-\frac{m_{Q}^{2}}{\alpha M^{2}}\bigg)
\frac{1}{\alpha^{2} M^{2}}e^{-\frac{m_{Q}^{2}}{\alpha M^{2}}},\nonumber\\
\hat{B}\Pi_{II}^{\langle\bar{q}q\rangle\langle g^{2}G^{2}\rangle^{2}}&=&\frac{m_{Q}\langle\bar{q}q\rangle\langle g^{2}G^{2}\rangle^{2}}{3^{4}\cdot2^{11}\pi^{4}}\int_{0}^{1}d\alpha
\bigg(3-\frac{m_{Q}^{2}}{\alpha M^{2}}\bigg)
\frac{1}{\alpha^{2} M^{2}}e^{-\frac{m_{Q}^{2}}{\alpha M^{2}}},\nonumber\\
\hat{B}\Pi_{III}^{\langle\bar{q}q\rangle\langle g^{2}G^{2}\rangle^{2}}&=&-\frac{m_{Q}\langle\bar{q}q\rangle\langle g^{2}G^{2}\rangle^{2}}{3^{4}\cdot2^{10}\pi^{4}}\int_{0}^{1}d\alpha
\bigg(3-\frac{m_{Q}^{2}}{\alpha M^{2}}\bigg)
\frac{1}{\alpha^{2} M^{2}}e^{-\frac{m_{Q}^{2}}{\alpha M^{2}}},\nonumber\\
\hat{B}\Pi_{IV}^{\langle\bar{q}q\rangle\langle g^{2}G^{2}\rangle^{2}}&=&\frac{m_{Q}\langle\bar{q}q\rangle\langle g^{2}G^{2}\rangle^{2}}{3^{4}\cdot2^{10}\pi^{4}}\int_{0}^{1}d\alpha
\bigg(3-\frac{m_{Q}^{2}}{\alpha M^{2}}\bigg)
\frac{1}{\alpha^{2} M^{2}}e^{-\frac{m_{Q}^{2}}{\alpha M^{2}}},\nonumber\\
\hat{B}\Pi_{I}^{\langle\bar{q}q\rangle^{2}\langle g^{2}G^{2}\rangle}&=&\frac{\langle\bar{q}q\rangle\langle g^{2}G^{2}\rangle}{3^{3}\cdot2^{7}\pi^{2}}
\bigg[4\langle\bar{s}s\rangle\bigg(1+\frac{m_{Q}^{2}}{M^{2}}\bigg)+m_{s}m_{Q}^{3}\Big(4\langle\bar{q}q\rangle-\langle\bar{s}s\rangle\Big)\frac{1}{(M^{2})^{2}}\bigg]
e^{-\frac{m_{Q}^{2}}{M^{2}}}\nonumber\\
&+&\frac{m_{Q}\langle\bar{q}q\rangle\langle g^{2}G^{2}\rangle}{3^{3}\cdot2^{6}\pi^{2}}\int_{0}^{1}
\frac{d\alpha}{\alpha^{2}M^{2}}\bigg[\Big(-2m_{Q}\langle\bar{s}s\rangle-3m_{s}\langle\bar{s}s\rangle+6m_{s}\langle\bar{q}q\rangle\Big)\nonumber\\
&+&m_{s}m_{Q}^{2}\Big(\langle\bar{s}s\rangle-2\langle\bar{q}q\rangle\Big)\frac{1}{\alpha M^{2}}\bigg]e^{-\frac{m_{Q}^{2}}{\alpha M^{2}}},\nonumber\\
\hat{B}\Pi_{II}^{\langle\bar{q}q\rangle^{2}\langle g^{2}G^{2}\rangle}&=&\frac{\langle\bar{q}q\rangle\langle g^{2}G^{2}\rangle}{3^{3}\cdot2^{7}\pi^{2}}
\bigg[-4\langle\bar{s}s\rangle\bigg(1+\frac{m_{Q}^{2}}{M^{2}}\bigg)+m_{s}m_{Q}^{3}\Big(4\langle\bar{q}q\rangle+\langle\bar{s}s\rangle\Big)\frac{1}{(M^{2})^{2}}\bigg]
e^{-\frac{m_{Q}^{2}}{M^{2}}}\nonumber\\
&+&\frac{m_{Q}\langle\bar{q}q\rangle\langle g^{2}G^{2}\rangle}{3^{3}\cdot2^{6}\pi^{2}}\int_{0}^{1}
\frac{d\alpha}{\alpha^{2}M^{2}}\bigg[\Big(2m_{Q}\langle\bar{s}s\rangle+3m_{s}\langle\bar{s}s\rangle+6m_{s}\langle\bar{q}q\rangle\Big)\nonumber\\
&-&m_{s}m_{Q}^{2}\Big(\langle\bar{s}s\rangle+2\langle\bar{q}q\rangle\Big)\frac{1}{\alpha M^{2}}\bigg]e^{-\frac{m_{Q}^{2}}{\alpha M^{2}}},\nonumber\\
\hat{B}\Pi_{III}^{\langle\bar{q}q\rangle^{2}\langle g^{2}G^{2}\rangle}&=&\frac{\langle\bar{q}q\rangle\langle g^{2}G^{2}\rangle}{3^{3}\cdot2^{6}\pi^{2}}
\bigg[4\langle\bar{s}s\rangle\bigg(1+\frac{m_{Q}^{2}}{M^{2}}\bigg)+m_{s}m_{Q}^{3}\Big(8\langle\bar{q}q\rangle-\langle\bar{s}s\rangle\Big)\frac{1}{(M^{2})^{2}}\bigg]
e^{-\frac{m_{Q}^{2}}{M^{2}}}\nonumber\\
&+&\frac{m_{Q}\langle\bar{q}q\rangle\langle g^{2}G^{2}\rangle}{3^{3}\cdot2^{5}\pi^{2}}\int_{0}^{1}
\frac{d\alpha}{\alpha^{2}M^{2}}\bigg[\Big(-2m_{Q}\langle\bar{s}s\rangle-3m_{s}\langle\bar{s}s\rangle+12m_{s}\langle\bar{q}q\rangle\Big)\nonumber\\
&+&m_{s}m_{Q}^{2}\Big(\langle\bar{s}s\rangle-4\langle\bar{q}q\rangle\Big)\frac{1}{\alpha M^{2}}\bigg]e^{-\frac{m_{Q}^{2}}{\alpha M^{2}}},\nonumber\\
\hat{B}\Pi_{IV}^{\langle\bar{q}q\rangle^{2}\langle g^{2}G^{2}\rangle}&=&\frac{\langle\bar{q}q\rangle\langle g^{2}G^{2}\rangle}{3^{3}\cdot2^{6}\pi^{2}}
\bigg[-4\langle\bar{s}s\rangle\bigg(1+\frac{m_{Q}^{2}}{M^{2}}\bigg)+m_{s}m_{Q}^{3}\Big(8\langle\bar{q}q\rangle+\langle\bar{s}s\rangle\Big)\frac{1}{(M^{2})^{2}}\bigg]
e^{-\frac{m_{Q}^{2}}{M^{2}}}\nonumber\\
&+&\frac{m_{Q}\langle\bar{q}q\rangle\langle g^{2}G^{2}\rangle}{3^{3}\cdot2^{5}\pi^{2}}\int_{0}^{1}
\frac{d\alpha}{\alpha^{2}M^{2}}\bigg[\Big(2m_{Q}\langle\bar{s}s\rangle+3m_{s}\langle\bar{s}s\rangle+12m_{s}\langle\bar{q}q\rangle\Big)\nonumber\\
&-&m_{s}m_{Q}^{2}\Big(\langle\bar{s}s\rangle+4\langle\bar{q}q\rangle\Big)\frac{1}{\alpha M^{2}}\bigg]e^{-\frac{m_{Q}^{2}}{\alpha M^{2}}},\nonumber\\
\hat{B}\Pi_{I}^{\langle\bar{q}q\rangle^{2}\langle g^{3}G^{3}\rangle}&=&\frac{\langle\bar{q}q\rangle\langle g^{3}G^{3}\rangle}{3^{3}\cdot2^{7}\pi^{2}}
\int_{0}^{1}\frac{d\alpha}{\alpha^{2}M^{2}}\bigg[-\langle\bar{s}s\rangle+m_{Q}\Big(2m_{Q}\langle\bar{s}s\rangle+3m_{s}\langle\bar{s}s\rangle-6m_{s}\langle\bar{q}q\rangle\Big)
\frac{1}{\alpha M^{2}}\nonumber\\
&+&m_{s}m_{Q}^{3}\Big(-\langle\bar{s}s\rangle+2\langle\bar{q}q\rangle\Big)\frac{1}{\alpha^{2}(M^{2})^{2}}\bigg]e^{-\frac{m_{Q}^{2}}{\alpha M^{2}}},\nonumber\\
\hat{B}\Pi_{II}^{\langle\bar{q}q\rangle^{2}\langle g^{3}G^{3}\rangle}&=&\frac{\langle\bar{q}q\rangle\langle g^{3}G^{3}\rangle}{3^{3}\cdot2^{7}\pi^{2}}
\int_{0}^{1}\frac{d\alpha}{\alpha^{2}M^{2}}\bigg[\langle\bar{s}s\rangle-m_{Q}\Big(2m_{Q}\langle\bar{s}s\rangle+3m_{s}\langle\bar{s}s\rangle+6m_{s}\langle\bar{q}q\rangle\Big)
\frac{1}{\alpha M^{2}}\nonumber\\
&+&m_{s}m_{Q}^{3}\Big(\langle\bar{s}s\rangle+2\langle\bar{q}q\rangle\Big)\frac{1}{\alpha^{2}(M^{2})^{2}}\bigg]e^{-\frac{m_{Q}^{2}}{\alpha M^{2}}},\nonumber\\
\hat{B}\Pi_{III}^{\langle\bar{q}q\rangle^{2}\langle g^{3}G^{3}\rangle}&=&\frac{\langle\bar{q}q\rangle\langle g^{3}G^{3}\rangle}{3^{3}\cdot2^{6}\pi^{2}}
\int_{0}^{1}\frac{d\alpha}{\alpha^{2}M^{2}}\bigg[-\langle\bar{s}s\rangle+m_{Q}\Big(2m_{Q}\langle\bar{s}s\rangle+3m_{s}\langle\bar{s}s\rangle-12m_{s}\langle\bar{q}q\rangle\Big)
\frac{1}{\alpha M^{2}}\nonumber\\
&+&m_{s}m_{Q}^{3}\Big(-\langle\bar{s}s\rangle+4\langle\bar{q}q\rangle\Big)\frac{1}{\alpha^{2}(M^{2})^{2}}\bigg]e^{-\frac{m_{Q}^{2}}{\alpha M^{2}}},\nonumber\\
\hat{B}\Pi_{IV}^{\langle\bar{q}q\rangle^{2}\langle g^{3}G^{3}\rangle}&=&\frac{\langle\bar{q}q\rangle\langle g^{3}G^{3}\rangle}{3^{3}\cdot2^{6}\pi^{2}}
\int_{0}^{1}\frac{d\alpha}{\alpha^{2}M^{2}}\bigg[\langle\bar{s}s\rangle-m_{Q}\Big(2m_{Q}\langle\bar{s}s\rangle+3m_{s}\langle\bar{s}s\rangle+12m_{s}\langle\bar{q}q\rangle\Big)
\frac{1}{\alpha M^{2}}\nonumber\\
&+&m_{s}m_{Q}^{3}\Big(\langle\bar{s}s\rangle+4\langle\bar{q}q\rangle\Big)\frac{1}{\alpha^{2}(M^{2})^{2}}\bigg]e^{-\frac{m_{Q}^{2}}{\alpha M^{2}}},\nonumber\\
\hat{B}\Pi_{I}^{\langle\bar{q}q\rangle\langle g^{2}G^{2}\rangle\langle g\bar{q}\sigma\cdot G q\rangle}&=&
\frac{m_{Q}\langle g^{2}G^{2}\rangle}{3^{4}\cdot2^{8}\pi^{2}}\bigg[m_{Q}^{2}\Big(-3m_{Q}\langle\bar{q}q\rangle\langle g\bar{s}\sigma\cdot G s\rangle-3m_{Q}\langle\bar{s}s\rangle\langle g\bar{q}\sigma\cdot G q\rangle
+24m_{s}\langle\bar{q}q\rangle\langle g\bar{q}\sigma\cdot G q\rangle\nonumber\\
&-&4m_{s}\langle\bar{q}q\rangle\langle g\bar{s}\sigma\cdot G s\rangle-3m_{s}\langle\bar{s}s\rangle\langle g\bar{q}\sigma\cdot G q\rangle\Big)
\frac{1}{M^{2}}
-m_{s}m_{Q}^{4}\langle\bar{q}q\rangle\Big(6\langle g\bar{q}\sigma\cdot G q\rangle-
\langle g\bar{s}\sigma\cdot G s\rangle\Big)\frac{1}{(M^{2})^{2}}\nonumber\\
&+&3m_{s}\Big(2\langle\bar{q}q\rangle\langle g\bar{s}\sigma\cdot G s\rangle-12\langle\bar{q}q\rangle\langle g\bar{q}\sigma\cdot G q\rangle+3\langle\bar{s}s\rangle\langle g\bar{q}\sigma\cdot G q\rangle\Big)\bigg]
\frac{1}{(M^{2})^{2}}e^{-\frac{m_{Q}^{2}}{M^{2}}},\nonumber\\
\hat{B}\Pi_{II}^{\langle\bar{q}q\rangle\langle g^{2}G^{2}\rangle\langle g\bar{q}\sigma\cdot G q\rangle}&=&
\frac{m_{Q}\langle g^{2}G^{2}\rangle}{3^{4}\cdot2^{8}\pi^{2}}\bigg[m_{Q}^{2}\Big(3m_{Q}\langle\bar{q}q\rangle\langle g\bar{s}\sigma\cdot G s\rangle+3m_{Q}\langle\bar{s}s\rangle\langle g\bar{q}\sigma\cdot G q\rangle
+24m_{s}\langle\bar{q}q\rangle\langle g\bar{q}\sigma\cdot G q\rangle\nonumber\\
&+&4m_{s}\langle\bar{q}q\rangle\langle g\bar{s}\sigma\cdot G s\rangle+3m_{s}\langle\bar{s}s\rangle\langle g\bar{q}\sigma\cdot G q\rangle\Big)
\frac{1}{M^{2}}
-m_{s}m_{Q}^{4}\langle\bar{q}q\rangle\Big(6\langle g\bar{q}\sigma\cdot G q\rangle+
\langle g\bar{s}\sigma\cdot G s\rangle\Big)\frac{1}{(M^{2})^{2}}\nonumber\\
&-&3m_{s}\Big(2\langle\bar{q}q\rangle\langle g\bar{s}\sigma\cdot G s\rangle+12\langle\bar{q}q\rangle\langle g\bar{q}\sigma\cdot G q\rangle+3\langle\bar{s}s\rangle\langle g\bar{q}\sigma\cdot G q\rangle\Big)\bigg]
\frac{1}{(M^{2})^{2}}e^{-\frac{m_{Q}^{2}}{M^{2}}},\nonumber\\
\hat{B}\Pi_{III}^{\langle\bar{q}q\rangle\langle g^{2}G^{2}\rangle\langle g\bar{q}\sigma\cdot G q\rangle}&=&
\frac{m_{Q}\langle g^{2}G^{2}\rangle}{3^{4}\cdot2^{7}\pi^{2}}\bigg[m_{Q}^{2}\Big(-3m_{Q}\langle\bar{q}q\rangle\langle g\bar{s}\sigma\cdot G s\rangle-3m_{Q}\langle\bar{s}s\rangle\langle g\bar{q}\sigma\cdot G q\rangle
+48m_{s}\langle\bar{q}q\rangle\langle g\bar{q}\sigma\cdot G q\rangle\nonumber\\
&-&4m_{s}\langle\bar{q}q\rangle\langle g\bar{s}\sigma\cdot G s\rangle-3m_{s}\langle\bar{s}s\rangle\langle g\bar{q}\sigma\cdot G q\rangle\Big)
\frac{1}{M^{2}}
-m_{s}m_{Q}^{4}\langle\bar{q}q\rangle\Big(12\langle g\bar{q}\sigma\cdot G q\rangle-
\langle g\bar{s}\sigma\cdot G s\rangle\Big)\frac{1}{(M^{2})^{2}}\nonumber\\
&+&3m_{s}\Big(2\langle\bar{q}q\rangle\langle g\bar{s}\sigma\cdot G s\rangle-24\langle\bar{q}q\rangle\langle g\bar{q}\sigma\cdot G q\rangle+3\langle\bar{s}s\rangle\langle g\bar{q}\sigma\cdot G q\rangle\Big)\bigg]
\frac{1}{(M^{2})^{2}}e^{-\frac{m_{Q}^{2}}{M^{2}}},\nonumber\\
\hat{B}\Pi_{IV}^{\langle\bar{q}q\rangle\langle g^{2}G^{2}\rangle\langle g\bar{q}\sigma\cdot G q\rangle}&=&
\frac{m_{Q}\langle g^{2}G^{2}\rangle}{3^{4}\cdot2^{7}\pi^{2}}\bigg[m_{Q}^{2}\Big(3m_{Q}\langle\bar{q}q\rangle\langle g\bar{s}\sigma\cdot G s\rangle+3m_{Q}\langle\bar{s}s\rangle\langle g\bar{q}\sigma\cdot G q\rangle
+48m_{s}\langle\bar{q}q\rangle\langle g\bar{q}\sigma\cdot G q\rangle\nonumber\\
&+&4m_{s}\langle\bar{q}q\rangle\langle g\bar{s}\sigma\cdot G s\rangle+3m_{s}\langle\bar{s}s\rangle\langle g\bar{q}\sigma\cdot G q\rangle\Big)
\frac{1}{M^{2}}
-m_{s}m_{Q}^{4}\langle\bar{q}q\rangle\Big(12\langle g\bar{q}\sigma\cdot G q\rangle+
\langle g\bar{s}\sigma\cdot G s\rangle\Big)\frac{1}{(M^{2})^{2}}\nonumber\\
&-&3m_{s}\Big(2\langle\bar{q}q\rangle\langle g\bar{s}\sigma\cdot G s\rangle+24\langle\bar{q}q\rangle\langle g\bar{q}\sigma\cdot G q\rangle+3\langle\bar{s}s\rangle\langle g\bar{q}\sigma\cdot G q\rangle\Big)\bigg]
\frac{1}{(M^{2})^{2}}e^{-\frac{m_{Q}^{2}}{M^{2}}},\nonumber\\
\hat{B}\Pi_{I}^{\langle g^{3}G^{3}\rangle\langle g\bar{q}\sigma\cdot G q\rangle}&=&\frac{\langle g^{3}G^{3}\rangle}{3^{3}\cdot2^{11}\pi^{4}}
\int_{0}^{1}\frac{d\alpha}{\alpha^{2}M^{2}}\bigg[m_{s}\Big(\langle g\bar{s}\sigma\cdot G s\rangle-3\langle g\bar{q}\sigma\cdot G q\rangle\Big)\bigg(1-\frac{2m_{Q}^{2}}{\alpha M^{2}}\bigg)\nonumber\\
&-&6m_{Q}\langle g\bar{q}\sigma\cdot G q\rangle\bigg(3-\frac{m_{Q}^{2}}{\alpha M^{2}}\bigg)
\frac{1-\alpha}{\alpha}\bigg]e^{-\frac{m_{Q}^{2}}{\alpha M^{2}}},\nonumber\\
\hat{B}\Pi_{II}^{\langle g^{3}G^{3}\rangle\langle g\bar{q}\sigma\cdot G q\rangle}&=&\frac{\langle g^{3}G^{3}\rangle}{3^{3}\cdot2^{11}\pi^{4}}
\int_{0}^{1}\frac{d\alpha}{\alpha^{2}M^{2}}\bigg[m_{s}\Big(\langle g\bar{s}\sigma\cdot G s\rangle+3\langle g\bar{q}\sigma\cdot G q\rangle\Big)\bigg(1-\frac{2m_{Q}^{2}}{\alpha M^{2}}\bigg)\nonumber\\
&+&6m_{Q}\langle g\bar{q}\sigma\cdot G q\rangle\bigg(3-\frac{m_{Q}^{2}}{\alpha M^{2}}\bigg)
\frac{1-\alpha}{\alpha}\bigg]e^{-\frac{m_{Q}^{2}}{\alpha M^{2}}},\nonumber\\
\hat{B}\Pi_{III}^{\langle g^{3}G^{3}\rangle\langle g\bar{q}\sigma\cdot G q\rangle}&=&\frac{\langle g^{3}G^{3}\rangle}{3^{3}\cdot2^{9}\pi^{4}}
\int_{0}^{1}\frac{d\alpha}{\alpha^{2}M^{2}}\bigg[m_{s}\Big(\langle g\bar{s}\sigma\cdot G s\rangle-\frac{3}{2}\langle g\bar{q}\sigma\cdot G q\rangle\Big)\bigg(1-\frac{2m_{Q}^{2}}{\alpha M^{2}}\bigg)\nonumber\\
&-&3m_{Q}\langle g\bar{q}\sigma\cdot G q\rangle\bigg(3-\frac{m_{Q}^{2}}{\alpha M^{2}}\bigg)
\frac{1-\alpha}{\alpha}\bigg]e^{-\frac{m_{Q}^{2}}{\alpha M^{2}}},\nonumber
\end{eqnarray}
and
\begin{eqnarray}
\hat{B}\Pi_{IV}^{\langle g^{3}G^{3}\rangle\langle g\bar{q}\sigma\cdot G q\rangle}&=&\frac{\langle g^{3}G^{3}\rangle}{3^{3}\cdot2^{9}\pi^{4}}
\int_{0}^{1}\frac{d\alpha}{\alpha^{2}M^{2}}\bigg[m_{s}\Big(\langle g\bar{s}\sigma\cdot G s\rangle+\frac{3}{2}\langle g\bar{q}\sigma\cdot G q\rangle\Big)\bigg(1-\frac{2m_{Q}^{2}}{\alpha M^{2}}\bigg)\nonumber\\
&+&3m_{Q}\langle g\bar{q}\sigma\cdot G q\rangle\bigg(3-\frac{m_{Q}^{2}}{\alpha M^{2}}\bigg)
\frac{1-\alpha}{\alpha}\bigg]e^{-\frac{m_{Q}^{2}}{\alpha M^{2}}}.\nonumber
\end{eqnarray}

\end{appendix}


\clearpage
\end{document}